\documentstyle[psfig,conf_iap,10pt]{article}
\newcommand{\lya}{\mbox{Ly$\alpha$}}

\begin{document}
\heading{%
%
Structure in the \lya\ forest
%
} 
\par\medskip\noindent
\author{%
J. Liske${^1}$, J.K. Webb${^1}$
}
\address{%
Department of Astrophysics \& Optics, School of Physics,
University of New South Wales, Sydney 2052, Australia
}
%

\begin{abstract}
The spatial distribution of \lya\ forest absorption systems toward a
group of 8, closely spaced QSOs has been analysed and evidence for
large scale structure has been found at $\langle z \rangle = 2.8$.
Our technique is based on the first and second moments of the
transmission probability density function which is capable of
identifying and assessing the significance of regions of over- or
underdense \lya\ absorption. 
\linebreak The data has revealed at least two interesting
features. 1. An overdense structure at $z = 2.27$ which extends at
least over $\sim 8 \; h^{-1}$ comoving Mpc ($q_0 = 0.5$) in the plane
of the sky.  Metal absorption lines have been found at the same
redshift and thus a cluster or proto-cluster of galaxies seems to have
been discovered.  2. A void at $z = 2.97$, extending over $\sim 20 \;
h^{-1}$ comoving Mpc in the plane of the sky, possibly caused by a
locally increased UV ionising flux due to a foreground QSO.

\end{abstract}
\section{Introduction}
Recent work \cite{Lanzetta95} has shown that at least some fraction of
the \lya\ absorption lines seen in the spectra of low redshift QSOs
arises in the extended haloes of galaxies. At high redshift \lya\
absorbers are found to be strongly clustered \cite{Fernandez96} and
many contain ionised carbon \cite{Cowie95}. This suggests that some
fraction of high redshift \lya\ absorbers may also be identified with
the haloes of galaxies. Thus it seems likely that the \lya\ forest may
exhibit large scale structure.
 
However, it has been shown \cite{Fernandez96} that significant
clustering may be missed when using the classical tool of cluster
analysis, the two-point correlation function
\cite{Peebles,SYBT}. Here, we present a new technique to search for
non-randomness in the spatial distribution of the \lya\ forest based
on the first and second moments of the transmission probability
density function. This method is able to identify the strength, position
and scale of individual structures since it retains spatial
information.  It is fairly insensitive to noise and resolution
characteristics and is easy to apply in practice. The new technique
has been tested with the help of synthetic spectra and it was found to
be substantially more sensitive than a two-point correlation function
analysis.

\section{Results}

The method was applied to the spectra of a close group of eight QSOs
with a mean redshift of 2.97. The data \cite{Williger96} was kindly
made available to us by Gerry Williger (see also these proceedings).

\begin{figure}
\centerline{\vbox{
\psfig{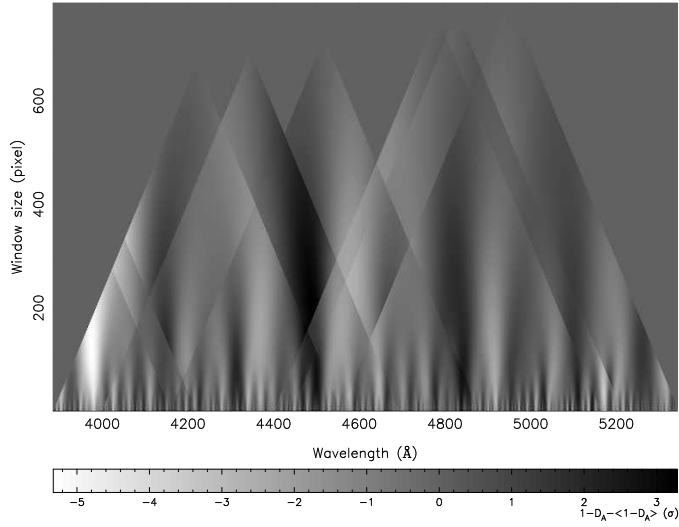}
}}
\caption[]{Transmission in the spectra of a group of eight, closely
spaced QSOs as a function of wavelength and size of window (smoothing)
function. The transmission is measured relative to a theoretical
expectation value and in units of the theoretical standard deviation
by convolving the spectra with a Gaussian of varying size. Each
``triangle'' corresponds to one spectrum, where the base is the
original spectrum itself and the tip is a value comparable to $1-D_A$,
where $D_A$ is the flux deficit parameter \cite{Oke82}.}
\label{liskeF1}
\end{figure}

Figure \ref{liskeF1} shows the result of the analysis. The most
prominent feature is a $5.3\sigma$ overdensity of absorption at
3978~\AA\ ($z = 2.272$). It is due to the spectra of Q0041-2707 and
Q0041-2658. The two lines of sight are separated by $2.4~h_{100}^{-1}$
proper Mpc ($q_0 = 0.5$) and the feature covers $\sim 2600$~km/s in
velocity space. Williger et~al.~\cite{Williger96} find metal absorption
at redshift 2.2722 in the spectrum of Q0041-2658, which is remarkably
consistent with the redshift of the overdense structure.

There are also two noticeable voids at $\sim 4490$~\AA\ and at
4842~\AA. The second void is possibly due to a foreground QSO which
lies within 500~km/s of the void.


%




%
\vfill
\end{document}